\documentclass[aps,prb,superscriptaddress,showpacs,floatfix,floats,nobalancelastpage,amsmath,amssymb,twocolumn]{revtex4-1}
\usepackage{graphicx}
\usepackage{color}
\usepackage{bbold}
\usepackage{amsfonts}
\usepackage{amsfonts, relsize}
\usepackage{latexsym,amsmath,amssymb,bm,euscript}
\usepackage{graphics}
\usepackage{graphicx}
\usepackage{hyperref}
\usepackage{color}
\usepackage{epstopdf}
\usepackage{subfigure}
\usepackage{bm}
\usepackage{bbm}
\usepackage{times}
\usepackage{multirow}
\usepackage{wasysym}
\usepackage{dcolumn}

\begin{document}

\title{Disordered and interacting parabolic semimetals in two and three dimensions}
\author{Hsin-Hua Lai}
\affiliation{National High Magnetic Field Laboratory, Florida State University, Tallahassee, Florida 32310, USA}

\author{Bitan Roy}
\affiliation{Condensed Matter Theory Center, University of Maryland, College Park, Maryland 20742-4111, USA}

\author{Pallab Goswami}
\affiliation{National High Magnetic Field Laboratory, Florida State University, Tallahassee, Florida 32310, USA}
\affiliation{Condensed Matter Theory Center, University of Maryland, College Park, Maryland 20742-4111, USA}

\date{\today}

\begin{abstract}
A clean noninteracting parabolic semimetal is characterized by quadratic band touching between the conduction and the valence bands at isolated diabolic points in the Brillouin zone and describes a fermionic quantum critical system with dynamic exponent z=2. We consider the stability of such a semimetal against electronic interaction and quenched disorder using a perturbative renormalization group analysis for two and three spatial dimensions. For the noninteracting problem infinitesimally weak disorder leads to an Anderson insulator and a diffusive metal respectively in two and three dimensions. On the other hand, the long range Coulomb interaction causes an excitonic instability for the clean interacting problem towards a broken symmetry ground state in both dimensions. Our weak coupling analysis of the combined effects of disorder and interaction suggests the competition between a broken symmetry and a disorder controlled metallic or insulating states, but is inadequate for describing the quantum phase transitions among them. We discuss the relevance of our results for bilayer graphene and some 227 iridate compounds, and identify these materials as promising candidates for exploring novel disorder and interaction controlled quantum critical phenomena.

\end{abstract}


\maketitle

\vspace{10pt}

\section{Introduction}

Recent years have witnessed a growing interest in semimetallic systems in both two (2D) and three (3D) spatial dimensions, which support gapless quasiparticle excitations only in the vicinity of isolated band touching points in the Brillouin zone (BZ). Around such a diabolic point located at $\mathbf{k}=\mathbf{K}_0$, the dispersion relations of the low energy quasiparticles become $\epsilon_{\pm}(\mathbf{k}) \sim |\mathbf{k}-\mathbf{K_0}|^z$, where $\pm$ respcetively describe the conduction and the valence bands. Consequently, the density of states (DOS) at low energies becomes $D(\epsilon) \sim |\epsilon|^{d/z-1}$, where $d$ is the spatial dimensionality and $z$ is the dynamic scaling exponent. When the Fermi energy is pinned to the band touching points, these semimetals can possess interesting power law behaviors for thermodynamic and transport quantities as a function of temperature or external frequency. These power laws as determined by $d$ and $z$ are generically different from the ones obtained for a standard Fermi liquid. In this respect, these semimetals provide striking examples of fermionic quantum criticality~\cite{Volovik, Ghosal, goswami-chakravarty}. Despite the difference in power law behaviors, these semimetals are not non-Fermi liquids, as the quasiparticle excitations in these systems are well defined, i. e., we have a finite quasiparticle residue.

There are many well known examples of the $z=1$ semimetals, which possess linearly dispersing, massless Dirac quasiparticles. Monolayer graphene~\cite{graphene-1, graphene-2, graphene-3} in 2D and Bi$_{1-x}$Sb$_{x}$~\cite{Lenoir,Ghosal}, Pb$_{1-x}$Sn$_{x}$Te~\cite{Dornhaus,goswami-chakravarty}, Cd$_3$As$_2$~\cite{weylexperiment1}, Na$_3$Bi~\cite{weylexperiment2} in 3D are some experimentally relevant examples of Dirac semimetal. Sometimes, these semimetals can also describe the universailty class of the quantum phase transition between two topologically distinct states. A 3D Dirac semimetal with an odd number of Dirac cones describes the quantum phase transition between a $Z_2$ topological insulator and a trivial band insulator~\cite{goswami-chakravarty}. By applying pressure or chemical doping Bi$_2$Se$_3$ can be tuned through a critical point that separates a strong $Z_2$ topological insulator from a trivial band insulator \cite{hassan-cava, ando, hassan-neupane, armitage, TPT-BiT}, where the quantum critical point can be described by one species of 3D massless Dirac fermion~\cite{zhang-model-PRB, goswami-chakravarty}. A similar transition in Bi$_{1-x}$Sb$_{x}$ produces three Dirac cones. In contrast, Pb$_{1-x}$Sn$_{x}$Te has four Dirac cones and describes the transition between a crystalline topological insulator and a trivial insulator~\cite{CYan}.

It is also conceivable to realize a $z=2$ parabolic semimetal for which the conduction and the valence bands show quadratic band touching. The prominent examples of such semimetals are the Bernel-stacked bilayer graphene~\cite{bilayer} in 2D, and HgTe~\cite{Dornhaus}, gray tin~\cite{alpha-tin-1}, and the normal state at high temperature for some 227 iridates such as Pr$_2$Ir$_2$O$_7$~\cite{Pr-1, Pr-2} (above $T \sim 1.5 K$) in 3D. While applied strain can covert HgTe into a strong topological insulator, stretching (if possible through negative chemical pressure) can induce a Dirac semimetal phase. The quantum phase transition between these two topologically distinct phases is succinctly described by the \emph{Luttinger model}~\cite{luttinger, murakami-zhang-nagaosa} which provides a simple realization of three dimensional, $z=2$ fermionic critical theory.

The stability of such noninteracting fermionic critical systems against disorder and interaction is a problem of deep fundamental importance. Dirac semimetals in both two and three spatial dimensions are extremely robust against weak electron-electron interaction, which is an \emph{irrelevant} perturbation due to the \emph{vanishing} DOS in the vicinity of the band touching point. This is captured by the scaling dimension of short range interaction $(z - d)$. Nevertheless, sufficiently strong interactions can drive Dirac semimetals through continuous phase transitions and take it into various broken symmetry phases, where the spectrum becomes fully gapped \cite{goswami-chakravarty, kveshchenko, HJR, drut, nandkishore-weyl, aji}. On the other hand, a long ago Abrikosov and Beneslavski noticed that Coulomb interaction is a \emph{relevant} perturbation in 3D parabolic semimetals with scaling dimension $z - 1 = 1$ and predicted the possibility of an infra-red stable non-Fermi liquid (NFL) fixed point in that system \cite{abrikosov-beneslavski, abrikosov}, which has recently been revisited by using the modern language of renormalization group (RG) and $\epsilon$-expansion \cite{balents-kim}. Besides the NFL phase, it has also been conjectured that parabolic semimetals instead may support various \emph{excitonic} ground states that in turn can replace the stable NFL phase at low temperatures \cite{serrington-kohn}. Interesting enough, a recent one-loop RG calculation clearly demonstrated that seemingly irrelevant short-range Coulomb interaction receives a strong boost from its long range tail and the NFL phase indeed becomes unstable towards the formation of a broken symmetry phase, even for arbitrarily weak bare strength of the short-ranged couplings \cite{herbut-jansen}. Taking into account only the short-range components of Coulomb interactions number of theoretical works have suggested the possibility of realizing numerous broken symmetry phases in bilayer grpahene \cite{vafek, nandkishore, aleiner, fanzhang, bitan-BLG}. Recent experiments are also strongly supportive of such phenomena in bilayer graphene at low temperatures, even in the absence of external magnetic or electric fields \cite{BLG-exp1, BLG-exp2, BLG-exp3, BLG-exp4}. One may therefore pose an interesting question: \emph{Analogous to 3D, is it possible to find a stable non-Fermi liquid phase in 2D parabolic semimetals
}? This is the question we address in this paper pursuing a one-loop RG appraoch. For the sake of technical ease, we here choose to work with a simpler model for spinless fermions in a 2D system that supports only one parabolic touching in BZ that, for example, can be realized in checkerboard or Kagome lattices~\cite{sun-fradkin-kivelson}. Nevertheless, our analysis can immediately be generalized for bilayer graphene.

The role of disorder in semimetallic systems is a subtle issue. Given that DOS scales as $D(\epsilon) \sim |\epsilon|^{d/z-1}$, the standard Born approximation dictates that quasiparticle lifetime ($\tau$) goes as $1/\tau \to 0$, when $z=2$ and $d > 2$. Therefore disorder appears to be an \emph{irrelevant} perturbation in 3D parabolic semimetals. However, the proper insight into the pertinence of disorder can only be gained within the framework of RG. The averaged disorder potential assumes the form of an four-fermion interaction term that is, however, \emph{nonlocal} (infinitely correlated) in imaginary time. The scaling dimension of the averaged disorder coupling ($\Delta$) is $[\Delta]=2z-d$. Therefore, scaling analysis dictates that any quenched short-ranged disorder is respectively a marginal and an irrelevant perturbation in 2D \cite{2d-diorderdirac} and 3D Dirac semimetals \cite{goswami-chakravarty, fradkin, roy-dassarma}. However, disorder is a \emph{relevant} perturbation in parabolic semimetals irrespective of its dimensionality, when $2 \leq d \leq 4$. In this paper we will perform disorder averaging using the \emph{replica formalism} and study its effects in parabolic semimetals in $d=2$ and $3$ through the procedure of coarse graining up to quadratic order in averaged disorder coupling ($\Delta$).

\begin{figure}[t]
\includegraphics[width=\columnwidth]{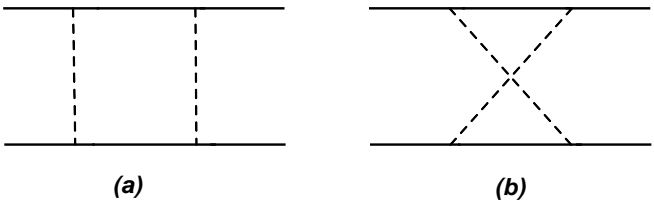}
\caption[] {A pair of one-loop diagrams, contributions from which do not cancel out. The dotted lines can represent disorder, long-range or short-range Coulomb interactions, which then generate different time-reversal symmetric disorder, density-density, or different  short-range interactions, respectively. Solid lines represent fermions.}
\label{lad-cross}
\end{figure}

Here we would like to point out a subtle intricacy involved in the RG procedure for parabolic semimetals. The noninteracting Hamiltonian $H_0(\vec{k})$ and consequently the Green's function $G^{-1}(i\omega, \vec{k})=i \omega-H_0 (\vec{k})$, where $\omega$ is the Matsubara frequency and $\vec{k}$ representing the spatial momenta, are defined in terms of $d$-wave or \emph{even-parity} functions. As a result, contributions form a pair of diagrams, shown in Fig.~\ref{lad-cross}, do not cancel each other since $G(-i \omega, -\vec{k})=G(-i\omega,\vec{k})$. This situation is in stark distinction from the one in Dirac semimetals, where the Green's function is defined in terms of $p$-wave or \emph{odd-parity} functions, and therefore $G(-i \omega, -\vec{k})=-G(i\omega,\vec{k})$. Consequently, a bare theory that contains only one interaction or one disorder, remains closed under coarse graining in Dirac semimetals. On the other hand, in parabolic semimetals even if the bare theory is defined only with the long-range Coulomb interaction or onsite potential disorder, it generates additional short range interaction and disorder vertices (time reversal symmetric), as we integrate out a shell of fast Fourier modes. In this paper, we will thoroughly discuss the importance of these diagrams and the couplings they generate in both 2D and 3D parabolic semimetals. Similar interplay between long range and short range interactions also arises for charged boson (Cooper pair)~\cite{RoyHerbutJuricic}  

We now provide a synopsis of our central results.
\begin{enumerate}

\item{ In this paper we first study the disorder effects on 2D parabolic semimetals. For simplicity, we consider a spinless model described by a two-component spinor with parabolic band touching. Through one loop RG calculations we show infinitesimally weak disorder is sufficient for destabilizing the parabolic semimetal. We argue that the disorder controlled phase is an Anderson insulator. These outcomes are germane to disordered bilayer graphene as well.}

\item{Quenched disorder remains a relevant perturbation in 3D parabolic semimetals as well. Infinitesimally weak disorder drives the parabolic semimetal into a diffusive metallic phase, where the DOS, the quasiparticle lifetime, and the mean-free path all defined at zero energy become finite, before it turns into an Anderson insulator at sufficiently large disorder.}

\item{We also study the effect of electronic correlations in clean parabolic semimetals in two and three spatial dimensions. In 2D, the short range interaction is a marginally relevant perturbation and its flow under RG suggests an instability of this system towards the formation of topological quantum anomalous Hall insulator at lowest temperature. Otherwise, the long range tail of Coulomb interactions enhances such ordering tandency irrespective of the sign of the bare short range interaction. The long range Coulomb interaction also leads to non mean-field corrections to various physical observables, such as critical temperature, scaling of specific heat. }

\item{In 3D the situation is more interesting. If we neglect the long range piece of Coulomb interactions, the RG flow equations can be closed in terms of \emph{two} linearly independent short range quartic couplings. The set of coupled RG equations displays a \emph{critical point}, which possibly indicates the appearance of a broken symmetry phase that breaks the \emph{rotational invariance} of the system in the ordered phase beyond a critical threshold of interaction strength. On the other hand, when the long range part of Coulomb interactions is restored such critical point gets vanquished and arbitrarily weak short-range interactions run away toward strong couplings. Therefore, clean interacting 3D parabolic semimetal is also expected to discover itself in a strongly coupled broken symmetry phase, as our one loop RG calculations would suggest \cite{herbut-jansen}.}

\item{Finally we address the interplay of interaction and disorder in 2D and 3D parabolic semimetals. Unfortunately both perturbations are relevant at the noninteracting fixed point and RG flows cannot be controlled as all the couplings grow stronger during the procedure of coarse graining. Nevertheless, from the flow equations of disorder and interaction couplings we can roughly estimate a qualitative  phase boundary between a broken symmetry phase and Anderson insulator in 2D or diffusive metal in 3D systems.}

\end{enumerate}

Rest of the paper is organized as follows. In the next section, Sec.~\ref{Sec:II}, we discuss the effect of short ranged quenched disorder in 2D parabolic semimetals. Sec. \ref{Sec:III} is devoted to discussing similar issues in three dimensional systems. Clean interacting two dimensional parabolic semimetal will be the subject of discussion in Sec. \ref{Sec:IV}, and the role of electronic correlations in three-dimensional systems are addressed in Sec. \ref{Sec:V}. Interplay of disorder and interactions in parabolic semimetals are discussed in Sec. VI in 2D and 3D. We summarize our findings and discuss the experimental relevance in Sec. \ref{Sec:VII}.

\section{Two dimensional non-interacting model with disorder}\label{Sec:II}

We begin with the continuum limit of the tight binding model for spinless fermions on a checkerboard lattice, introduced by Sun {\it et al.} \cite{sun-fradkin-kivelson}, which provides the simplest example of the parabolic band touching in two dimensions at the $\Gamma$ point, $\mathbf{K}_0=0$. The effective action for the low energy quasiparticles is
\begin{equation}
S^{2D}_0=\int d^2x d\tau \; \psi^\dagger(x,\tau) \; [\partial_\tau + \mathbf{d} \cdot \boldsymbol \sigma] \; \psi(x,\tau),\label{2Dfree}
\end{equation}
where $\psi(x)$ is a two-component spinor (which arises from the two-atom unit cell), and
\begin{equation}
\mathbf{d}=(d_1,d_2,d_3)=\left(-\frac{\partial^2_x-\partial^2_y}{m},0,-\frac{2\partial_x\partial_y}{m}\right).
\end{equation}
In the above equation $m$ is the effective mass of the quasiparticles, and the spatial integrals have a lower cut-off $\Lambda^{-1} \sim a$, where $a$ is the lattice spacing, and we have set $\hbar =1$. For convenience, we perform a unitary transformation $\psi \to \exp(i \pi \sigma_1/4) \psi$, under which the $\mathbf{d} \to (d_1,d_3,d_2)$, and the fermion propagator becomes
\begin{equation}
G(i\omega,\mathbf{k})=-\frac{i\omega+d_1\sigma_1+d_3\sigma_2}{\omega^2+\mathbf{d}^2_{\mathbf{k}}}.
\end{equation}

Now we introduce the following time reversal symmetric disorder bilinears
\begin{equation}
S^{2D}_d= \int d^2x  d\tau \; \psi^\dagger(x,\tau) \; \left [V_0(x) \mathbb{1}_2 + V_a(x) \sigma_a \right]\; \psi(x,\tau),\label{2Ddis}
\end{equation}
 where $V_0(x)$ represents the random chemical potential, $V_{a=1,2} (x)$ arises from random hopping, and the repeated index $a$ represents the summation over $a=1,2$. The total effective action in the presence of disorder is $S^{2D}=S^{2D}_0+S^{2D}_d$. It will become clear from our RG analysis that a random hopping gets generated during the process of coarse graining, even if we start only with a random chemical potential. Therefore, it is important to treat both types of disorder on equal footing. For simplicity we choose Gaussian white noise distribution for the disorder potentials according to
\begin{eqnarray}
\langle \langle V_0(x) V_0(x^\prime) \rangle \rangle &=&\frac{2\pi \Lambda^2 \Delta_1}{m^2} \; \delta^2(x-x^\prime)\\
\langle \langle V_a(x) V_a(x^\prime) \rangle \rangle &=& \frac{2\pi \Lambda^2 \Delta_2}{m^2}\;  \delta^2(x-x^\prime),
\end{eqnarray}
where $\Delta_1$ and $\Delta_2$ are two dimensionless disorder coupling constants. We perform the average over disorder by employing replica method. The replicated disorder averaged action is given by
\begin{eqnarray}
\bar{S}^{2D}&=& \int d^2x d\tau \psi^\dagger_a(x,\tau) \; [\partial_\tau + \mathbf{d} \cdot \boldsymbol \sigma] \; \psi_a(x,\tau) \nonumber \\
&-&\frac{\pi \Lambda^2 \Delta_1}{m^2} \sum_{j=1}^{2}\int d^2x d\tau d\tau^\prime (\psi^\dagger_a \sigma_j\psi_a)_{(x,\tau)}(\psi^\dagger_b \sigma_j\psi_b)_{(x,\tau^\prime)} \nonumber \\
&-& \frac{\pi \Lambda^2 \Delta_2}{m^2} \int d^2x d\tau d\tau^\prime (\psi^\dagger_a\psi_a)_{(x,\tau)}(\psi^\dagger_b\psi_b)_{(x,\tau^\prime)}
\end{eqnarray}
Now we expand the disorder induced four fermi interaction terms up to the second order in the disorder couplings, and perform the perturbative RG calculations using the Wilsonian momentum shell scheme. We integrate out the fast degrees of freedom from the shell $\Lambda e^{-l} < k < \Lambda$, and obtain
\begin{eqnarray}
&&\bar{S}^{2D,\prime}= \int d^2x d\tau \psi^\dagger_a(x,\tau) \; [\partial_\tau(1+Al) + \mathbf{d} \cdot \boldsymbol \sigma] \; \psi_a(x,\tau) \nonumber \\
&-&\frac{\pi \Lambda^2 \Delta_1}{m^2}(1+Bl) \int d^2x d\tau d\tau^\prime (\psi^\dagger_a\psi_a)_{(x,\tau)}(\psi^\dagger_b\psi_b)_{(x,\tau^\prime)} \nonumber \\
&-& \frac{\pi \Lambda^2 \Delta_2}{m^2}(1+Cl) \sum_{j=1}^{2}\int d^2x d\tau d\tau^\prime (\psi^\dagger_a \sigma_j\psi_a)_{(x,\tau)} \nonumber \\
&\times& (\psi^\dagger_b \sigma_j\psi_b)_{(x,\tau^\prime)},
\end{eqnarray}
where $A=\Delta_1+2\Delta_2$, $B=2\Delta_1+8\Delta_2$, and $C=4\Delta_2+\Delta^2_1/(2\Delta_2)$. After rescaling the space-time coordinates according to $x \to x e^l$, $\tau \to \tau e^{z(l)l}$, where $z(l)$ is a scale dependent dynamic exponent, and $\psi \to \psi Z_{\psi}^{1/2}$, with the field renormalization constant
\begin{equation}
Z_\psi=e^{2l}[1+(\Delta_1+2\Delta_2)l],
\end{equation}
we bring $\bar{S}^{2D,\prime}$ back to the original form of $\bar{S}^{2D}$. Through this procedure we find the following RG flow equations
\begin{eqnarray}
\frac{d \log m}{dl}&=&-(z-2-\Delta_1-2\Delta_2), \\
\frac{d\Delta_1}{dl}&=&2\Delta_1(1+\Delta_1+4\Delta_2),\\
\frac{d\Delta_2}{dl}&=&2\Delta_2(1+2\Delta_2)+\Delta^2_1.
\end{eqnarray} For extracting physically meaningful information, we hold $m$ fixed under the RG flow, which dictates the following formula for the dynamic exponent
\begin{equation}
z(l)=2+\Delta_1+2\Delta_2.
\end{equation}
These RG equations have an unstable fixed point at $\Delta_1=\Delta_2=0$, with $z=2$, which describes the clean, noninteracting parabolic band touching problem. Both types of disorder flow to strong coupling regime. Hence, infinitesimally weak disorder is sufficient to destroy the ballistic, parabolic semimetal in 2D. Notice, that even if we choose the bare coupling $\Delta_2=0$, it will be generated in the process of coarse graining.

The perturbative RG calculations by itself can not reveal the nature of the disorder dominated phase. For a better understanding of this phase we can perform a self-consistent Born approximation calculation, which is now justified by the strong coupling flow of disorder. Due to the finite density of states at zero energy, even the simple Born approximation shows that an infinitesimally weak disorder is sufficient to cause a finite lifetime $\tau \sim m \Lambda^{-2} \Delta^{-1}$ for the quasiparticle, where $\Delta=\Delta_1+2\Delta_2$. This also implies the existence of a mean free path $\ell \sim \Lambda^{-1} \Delta^{-1/2}$. By solving the RG equations one can obtain a better estimate for $\tau$ and $\ell $. For length scales bigger than $\ell$ the ballistic quasiparticle picture breaks down, and a diffusive metal phase will be the suitable starting point, which can be described in terms of a non-linear sigma model. At scales much larger than $\ell$, system behaves as an Anderson insulator.

In terms of the non-linear sigma model description for the diffusion modes, there is a sharp distinction between the strong disorder regime of dirty two-component Dirac and parabolic fermions. The non-linear sigma model for the Dirac fermion allows a $Z_2$ topological $\theta$-term, where $\theta=N_v \pi$, and $N_v=\pm 1$ is the vorticity of the Bloch bands~\cite{Mirlin_nonlinearsigma,Ryu_nonlinearsigma}. Due to the existence of this $\theta=\pi \; \mathrm{mod} \; 2\pi$, the sigma model remains gapless and the two-component Dirac fermion evades Anderson localization. In contrast the parabolic touching corresponds to $N_v=\pm 2$, and the theta term becomes $0 \; \mathrm{mod} \; 2\pi$. For this reason, the two dimensional non-linear sigma model for the parabolic fermions can not remain gapless, and Anderson localization becomes inevitable.

\section{Three dimensional non-interacting model with disorder}\label{Sec:III}

The parabolic band touching in 3D can be well described by the Luttinger Hamiltonian \cite{luttinger, murakami-zhang-nagaosa}. Now the effective action for the low energy quasiparticles is given by
\begin{equation}
S^{3D}_0=\int d^3x d\tau \; \psi^\dagger(x,\tau) \; [\partial_\tau + \mathbf{d} \cdot \boldsymbol \Gamma] \; \psi(x,\tau),\label{3Dfree}
\end{equation}
where $\psi(x)$ is a four-component spinor, and the five components of the vector
\begin{eqnarray}
\mathbf{d}&=&(d_1,d_2,d_3,d_4,d_5)=\frac{1}{m}\bigg(\sqrt{3}\partial_y\partial_z,\sqrt{3}\partial_z\partial_x,\sqrt{3}\partial_x\partial_y, \nonumber \\
&& \frac{\sqrt{3}}{2}(\partial^2_x-\partial^2_y),\frac{1}{2}(\partial^2_x+\partial^2_y-2\partial^2_z)\bigg)
\end{eqnarray}
correspond to five l=2 or $d$-wave spherical harmonics. The five mutually anticommuting $\Gamma$ matrices are defined according to $\Gamma_1=\sigma_3 \otimes \tau_2$, $\Gamma_2=\sigma_3 \otimes \tau_1$, $\Gamma_3=\sigma_2 \otimes \tau_0$, $\Gamma_4=\sigma_1 \otimes \tau_0$, and $\Gamma_5=\sigma_3 \otimes \tau_3$, where $\tau_0$ is $2 \times 2$ identity matrix \cite{murakami-zhang-nagaosa}.

For this 3D model, we can consider the following time reversal symmetric disorder bilinears
\begin{equation}
S^{3D}_d= \int d^3x  d\tau \; \psi^\dagger(x,\tau) \; \left [V_0(x) \mathbb{1}_4 + V_a(x) \Gamma_a \right]\; \psi(x,\tau),\label{3Ddis}
\end{equation}
where $V_0$ is a random chemical potential, $V_a(x)$ arises from the random spin-orbit scattering, and repeated index $a$ represents the summation over $a=1,2,3,4,5$. As in the 2D problem, even if we set the bare strength of $V_a$ to be zero, we generate this type of disorder during the procedure of coarse graining. We again choose Gaussian white noise distribution for the random chemical potential and spin-orbit coupling according to
\begin{eqnarray}
\langle \langle V_0(x) V_0(x^\prime) \rangle \rangle &=&\frac{2\pi^2 \Lambda\Delta_1}{m^2} \; \delta^3(x-x^\prime)\\
\langle \langle V_a(x) V_a(x^\prime) \rangle \rangle &=&\frac{2\pi^2 \Lambda\Delta_2}{m^2} \;  \delta^3(x-x^\prime),
\end{eqnarray}
respectively, where $\Delta_1$ and $\Delta_2$ are two dimensionless disorder coupling constants. Now following the same procedure of replica average and subsequent coarse graining using the momentum shell method, we arrive at the following RG flow equations
\begin{eqnarray}
\frac{d \log m}{dl}&=&-(z-2-\Delta_1-5\Delta_2), \\
\frac{d\Delta_1}{dl}&=&\Delta_1(1+2\Delta_1+14\Delta_2),\\
\frac{d\Delta_2}{dl}&=&\Delta_2 \left(1+\frac{32}{5}\Delta_2-\frac{6}{5}\Delta_1\right)+\frac{2}{5}\Delta^2_1.
\end{eqnarray}
When we hold $m$ fixed, we find
\begin{equation}
z(l)=2+\Delta_1+5\Delta_2.
\end{equation}
Again, the parabolic semimetal fixed point at $\Delta_1=\Delta_2=0$ with $z=2$ is unstable against disorder.

For a better understanding of the disorder controlled phase we can perform a self-consistent Born approximation calculation. The naive Born approximation which suggests zero scattering rate at zero energy is clearly inadequate for this purpose. Even though the DOS at zero energy for 3D parabolic semimetal vanishes as $\sqrt{E}$, the disorder couplings have positive scaling dimension \emph{one}, which gives rise to the mean free path $\ell \sim \Lambda^{-1} \Delta^{-1}$ and the life-time  $\tau \sim \ell^2$ at zero energy. This corresponds to the destabilization of the ballistic, incompressible semimetal phase in favor of a compressible, diffusive metal. This phase can again be described in terms of the non-linear sigma model for the diffusion modes. Unlike in two dimensions, the diffusive metal can remain stable up to a certain strength of disorder couplings before succumbing to the Anderson insulator.

\section{Two dimensional clean interacting model}\label{Sec:IV}

Before discussing the more realistic problem of the combined effects of interaction and disorder, we analyze the clean, interacting model to understand the role of electronic correlations. For clarity, we will consider two and three spatial dimensions separately. The 2D model only in the presence of the short range interactions has been considered in literature \cite{sun-fradkin-kivelson}, where it has been shown that the short range interaction is a marginally relevant perturbation, which causes a BCS like instability towards an insulating, anomalous quantum Hall state at zero temperature. Similar mechanism also remains operative in bilayer graphene which can support a plethora of broken symmetry phases due to the presence of additional flavor degrees of freedom~\cite{vafek, nandkishore, aleiner, fanzhang}.

Thus far the long range Coulomb interaction and the short range interactions have not been treated on equal footing. The long range Coulomb interaction at tree level is a relevant perturbation with scaling dimension $z-1=1$, and can have more dominant effects than the short range interaction. It is also legitimate to ask, whether the Coulomb interaction after many body screening effects can lead to an infra-red stable, NFL fixed point as discussed by Abrikosov and Beneslavskii for the 3D parabolic semimetal~\cite{abrikosov-beneslavski, abrikosov}. So far, RPA/ large N technique has been adopted by several authors~\cite{nandkishore,aleiner,Barlas}, which produces static screening of the Coulomb interaction due to the finite DOS at zero energy. Since, in this process one first integrates out all the states from zero energy up to the ultra-violet cutoff, a subsequent coarse graining always remains a questionable procedure. For this reason we will perform a weak coupling RG analysis and treat both short range and long range interactions on an equal footing. In this and the subsequent section we will show a remarkable similarity for the flow of the Coulomb coupling both in 2D and 3D, and its strong feedback effects to the coupling constants for the short range interactions.

The interacting part of the effective action is given by
\begin{eqnarray}
S_{int,2D}&=&g \int d^2x \; d\tau (\psi^{\dagger}\sigma_3\psi)^2 + \frac{e^2}{2} \int d^2x \int d^2x^\prime \nonumber \\
&\times& \int d\tau (\psi^{\dagger}\psi)_{x,\tau} \frac{1}{|\mathbf{x}-\mathbf{x}^\prime|}(\psi^{\dagger}\psi)_{x^\prime,\tau},
\end{eqnarray}
where $\psi$ is a two-component spinor and the non-interacting action is given in Eq.~(\ref{2Dfree}). We carry out the perturbative RG calculations for $S^{2D}_0+S_{int,2D}$ to the quadratic order in terms of the dimensionless couplings $g \to gm/(4\pi)$ and $\alpha=e^2m/(4\pi \Lambda)$. Our flow equations are
\begin{eqnarray}
\frac{d \log m}{dl}&=&-\left(z-2+\frac{3}{8}\alpha \right),\\
\frac{d \alpha}{dl}&=&(z-1)\alpha=\alpha \left(1-\frac{3}{8}\alpha\right), \\
\frac{dg}{dl}&=&g^2+\alpha^2+\frac{13}{8}g\alpha.
\end{eqnarray}
By holding $m$ fixed, we find a scale dependent dynamic exponent $z(l)=2-3\alpha(l)/8$. Notice, that in contrast to the disordered problem where the dynamic exponent is an increasing function, the Coulomb interaction rather causes a decrease of $z$. Physically this can be understood in the following way. Disorder coupling tends to enhance the DOS and lead to a slower dynamics, which leads to an increment of $z$. On the other hand, the Coulomb interaction causes the depletion of the DOS at low energy, while causing a spectral shift to higher energy, resulting in a faster dynamics or decrease of $z$.

The above equations show that the parabolic semimetal corresponds to an unstable fixed point $\alpha=g=0$, $z=2$, and the short range interaction flows towards $+\infty$, irrespective of its bare value and sign. Even if we started with the bare value of the short range coupling $g=0$, it gets generated in the process of coarse graining, which is a consequence of the combined effects of two diagrams shown in Fig. 1, where the dashed lines represent Coulomb interaction. The Coulomb coupling constant $\alpha$ has the tendency of flowing towards $\alpha_c=8/3$, with $z=1$. If we have not treated the short range interaction on an equal footing $\alpha_c$ would have emerged as an infrared-stable NFL fixed point. This is the two dimensional analog of the NFL fixed point, discussed by Abrikosov-Beneslavskii in 3D. However, our treatment of both short and long range interactions shows that this NFL fixed point is unstable towards a broken symmetry phase at low temperatures (due to the strong relevance of $g$).

The equation for $\alpha(l)$ can be solved analytically giving
\begin{equation}
\alpha(l)=\frac{\alpha_c}{1+\left ( \frac{\alpha_c}{\alpha_0}-1\right)e^{-l}},
\end{equation}
where $\alpha_0=\alpha(l=0)$ is the bare value of Coulomb coupling. However, for a generic value of $\alpha$ the equation for $g(l)$ needs to be solved numerically. From the solution of $g(l)$ we can estimate the length or the energy scale for the onset of the broken symmetry phase. For the particular value of $\alpha=\alpha_c$, we can solve the equation for $g(l)$ analytically to obtain
\begin{equation}
g(l)=-x\alpha_c+y \alpha_c \tan \left[ \arctan \left( \frac{g(0)+x \alpha_c}{y\alpha_c} \right)+y \alpha_c l\right],
\end{equation}
where $x=13/16$ and $y=\sqrt{87}/16$. The scale ($l^\ast$) where perturbative RG breaks down can be obtained either by setting $g(l^\ast) \sim 1$ or $g(l^\ast) \to \infty$, and this roughly corresponds to the onset of the broken symmetry. The estimated scales from these two conventions only differ through non-universal numerical factors. If we choose $g(0)=0$ and $g(l^\ast) \to \infty$, we end up with
\begin{equation}
l^\ast=\frac{1}{y\alpha_c}\left(\frac{\pi}{2}-\arctan \frac{x}{y} \right) \sim \frac{1}{\alpha_c},
\end{equation} which gives rise to a very large transition temperature
\begin{equation}
T_c \sim T_0 e^{-1/\alpha_c},
\end{equation}
where $T_0= \Lambda^2/m$ is the ultra-violet cutoff. For the noninteracting model, the scaling behaviors of various physical quantities are determined by $d=2$, $z=2$. However, the flow of Coulomb coupling towards $\alpha_c$ causes corrections to the scaling properties. Only for the weak coupling (small $\alpha$ and $g$) problem, $l^\ast$ can be considerably large and the NFL behavior can be observed over a significant range of temperature.

\section{Three dimensional clean interacting model}\label{Sec:V}

Following the same strategy as in the 2D problem, we can discuss the RG analysis of the 3D clean interacting model. The noninteracting action for this problem is given in Eq.~(\ref{3Dfree}). The underlying rotational symmetry of the system dictates that there are all together three types of short-ranged interactions of the form $(\psi^\dagger \psi)^2$, $(\psi^\dagger\Gamma_j \psi)^2$ and $(\psi^\dagger \Gamma_{ij} \psi)^2$, where $\Gamma_{ij}=i \Gamma_i \Gamma_j$ and summation over the indices $i,j=1,2,\cdots, 5$ is assumed but with a restriction $j>i$.
However, all three quartic couplings are not linearly independent and there exists a mathematical constraint, known as the \emph{Fierz identity}, that allows to write the last four-fermion term as a linear combination of the remaining two according to~\cite{HJR}
\begin{equation}\label{fierz3d}
10(\psi^\dagger \psi)^2+2(\psi^\dagger \Gamma_j \psi)^2=(\psi^\dagger \Gamma_{ij} \psi)^2.
\end{equation}
A detailed derivation of this algebraic identity can be found in Appendix A. For this reason we begin with the following effective action for the interacting part
\begin{eqnarray}
&&S^{3D}_{int}= \int d^3x \; d\tau \left[g_1(\psi^{\dagger}\psi)^2+g_2(\psi^\dagger \Gamma_j \psi)^2\right]+\nonumber \\
&&+ \frac{e^2}{2} \int d^3x \int d^3x^\prime \int d\tau (\psi^{\dagger}\psi)_{x,\tau} \frac{1}{|\mathbf{x}-\mathbf{x}^\prime|}(\psi^{\dagger}\psi)_{x^\prime,\tau}. \nonumber \\
\end{eqnarray}
During the process of course graining if we generate the four-fermion term of the form $(\psi^\dagger \Gamma_{ij} \psi)^2$, we rewrite it back in terms of the couplings $g_1$ and $g_2$ according to Eq.~(\ref{fierz3d}). Since, the Fourier transform of the Coulomb potential in 3D is $1/q^2$, electric charge receives a direct screening correction from the particle-hole bubble. The particle-hole excitations also cause a screening correction of the short range interaction in the charge channel (denoted by coupling $g_1$) which is $\propto g_1 e^2$. These are the two new features of the diagrammatic perturbation theory of the 3D model and everything else is qualitatively similar to the 2D counterpart. The RG flow equations are given by (see also Ref.~\onlinecite{herbut-jansen})
\begin{eqnarray}
\frac{d \log m}{dl}&=&-\left(z-2+\frac{\alpha}{15} \right),\\
\frac{d \alpha}{dl}&=&\alpha \left(1-\frac{31}{15}\alpha\right), \\
\frac{dg_1}{dl}&=&g_1\left(z-3-\frac{1}{2}g_2-4\alpha \right)-\frac{5}{2}g^2_2-2\alpha g_2, \label{eq:g1}\\
\frac{dg_2}{dl}&=&g_2\left(z-3-\frac{63}{20}g_2+\frac{2}{5}g_1+\frac{8}{5}\alpha \right) \nonumber \\
&-&\frac{1}{20}g^2_1-\frac{4}{5}\alpha^2.
\end{eqnarray} Notice, that even if the bare value of $g_2$ is zero, it gets generated by the long range Coulomb coupling in the process of coarse graining from a pair of diagrams shown in Fig.~1 where the dashed lines represent Coulomb interaction \cite{herbut-jansen}. The only difference between our flow equations and the ones in Ref.~\onlinecite{herbut-jansen} is the $g_1\alpha$ term in Eq.~(\ref{eq:g1}), which describes the screening of the short range interaction due to the particle-hole bubble. This process only works in 3D, as the $q^2$ dependence of the bubble is compensated by the Coulomb propagator (goes as $1/q^2$). However, this correction is only quantitative in nature and does not alter the physical conclusions drawn in Ref.~\onlinecite{herbut-jansen}.

In the absence of the Coulomb coupling ($\alpha=0$), we have following two fixed points:
\begin{eqnarray}
&& (i) \; g_1=g_2=0, z=2 \\
&& (ii) \; g_1\simeq-0.398, \; g_2\simeq-0.361, \; z=2, 
\end{eqnarray} which respectively describe the stable parabolic semimetal phase, and a quantum critical point that governs the phase transition between the parabolic semimetal and a broken symmetry phase. By linearizing the flow equations around the quantum critical point we will find the correlation length exponent $\nu =1/(d-z)=1$. The correlation length exponent for a Gross-Neveu model being exactly equal to inverse of the bare scaling dimension of the short range coupling is an artifact of the lowest order RG calculations. However, the important point is to notice that we have an interacting, itinerant quantum critical point, which is usually hard to find in three dimensional models. In particular for $z=1$ problem of Dirac fermions, one sits at the upper critical dimension and finds mean-field critical behavior with $\nu=1/2$. For the $z=2$ Gross-Neveu problem at hand, the upper critical dimension is $d=4$. The susceptibility calculation suggests that the broken symmetry phase breaks rotational symmetry of the cubic lattice, but preserves the time-reversal symmetry. A simple consideration of energetics within the weak coupling approach suggests that $\psi^\dagger \Gamma_5 \psi (>0)$ is the most favorable order parameter, since it supports a fully gapped spectrum, and the ground state corresponds to a strong topological insulating phase~\cite{herbut-jansen}. The energy gap 
\begin{equation}
E_g \sim \delta^{\nu z} \sim \delta^2,
\end{equation} where $\delta$ is a function of $g_1$ and $g_2$, which is the relevant variable (obtained by linearizing the RG equations in the vicinity of the quantum critical point) and measures the deviation from the critical point. However, emergence of a time-reversal symmetry breaking magnetic order parameter ($\langle \psi^\dagger \Gamma_{45} \psi\rangle \neq 0$ for example) in the strong coupling regime can not be ruled out from this type of weak coupling calculations.

The long range Coulomb interaction dramatically alters this picture. By itself the Coulomb coupling $\alpha$ has the tendency to flow into the Abrikosov-Beneslavskii fixed point $\alpha_c=15/31$, $1<z<2$~ \cite{abrikosov-beneslavski}. However, the interplay between the short range and the long range interaction causes a runaway flow for the couplings $g_1$ and $g_2$, and no infra-red stable fixed point can be found within the weak coupling approach. In the analysis of Abrikosov-Beneslavskii \cite{abrikosov-beneslavski} and Moon {\it et al.} \cite{balents-kim} the self-consistent generation of short range coupling from the $\alpha$ has been ignored, which led to the infra-red stable, NFL fixed point. Our analysis suggests that there are NFL corrections to the scaling at high temperatures, which finally give away in favor of a broken symmetry phase at some critical/transition temperature. Ultimately, the weak coupling RG analysis of the long range interaction vindicates the existence of an excitonic condensate at low temperatures for both 2D and 3D clean interacting parabolic fermions, which is in agreement with the recent study by Herbut and Janssen \cite{herbut-jansen}.


\section{Combined effects of disorder and interaction}\label{Sec:VI}
Now we are in a position to consider the RG flow equations for the combined effects of disorder and interaction. The flow equations of the 2D model are now given by
\begin{eqnarray}
\frac{d \log m}{dl}&=&-\left(z-2+\frac{3}{8}\alpha -\Delta_1-2\Delta_2\right),\\
\frac{d \alpha}{dl}&=&(z-1)\alpha=\alpha \left(1-\frac{3}{8}\alpha+\Delta_1+2\Delta_2\right), \\
\frac{dg}{dl}&=&g^2+\alpha^2+\frac{13}{8}g\alpha+g(6\Delta_2-3\Delta_1),\\
\frac{d\Delta_1}{dl}&=&\Delta_1\left[2-\frac{3}{4}\alpha+2\Delta_1+8\Delta_2\right], \\
\frac{d \Delta_2}{dl}&=&\Delta_2\left[2+\frac{1}{4}\alpha +g+4\Delta_2\right]+\Delta^2_1.
\end{eqnarray} These equations show that both disorder and interaction run into strong coupling regime, and we can not find a quantum critical point. However from the numerical solution for the RG equations, we can identify two flow lengths $l_1$ and $l_2$, which respectively correspond to the length scales at which $g$ and the dominant disorder coupling become $\sim 1$. There is an interaction dominant region determined by $l_1<l_2$, when $g$ becomes $\sim1$ before the disorder coupling. In this region the system should enter the anomalous quantum Hall phase. Inside this phase there is a hard gap in the density of states, and the disorder induced scattering rate at zero energy indeed vanishes. For disorder dominated phase determined by $l_2<l_1$, we expect the system to be a topologically trivial Anderson insulator. Therefore, we expect a disorder and interaction controlled quantum phase transition between these  topologically distinct phases. In the language of non-linear sigma model, we anticipate the emergence of a $\theta=\pi$ term at the critical point to make the sigma model gapless~\cite{Pruisken}. 

For the 3D model the combined flow equations for disorder and interaction couplings are
\begin{eqnarray}
\frac{d \log m}{dl}&=&-\left(z-2+\frac{1}{15}\alpha -\Delta_1-5\Delta_2\right),\\
\frac{d\alpha}{dl}&=&\alpha\left[1-\frac{31}{15}\alpha+\Delta_1+5\Delta_2\right],\\
\frac{dg_1}{dl}&=&-g_1\left[1+\frac{g_2}{2}+\frac{61}{15}\alpha-\Delta_1-9\Delta_2\right] \nonumber \\
&&-\frac{5}{2}g^2_2+4\Delta_1g_2-2\alpha g_2, \\
\frac{dg_2}{dl}&=&-g_2\left[1-\frac{2}{5}g_1+\frac{63}{20}g_2+\frac{11}{5}\Delta_1-9\Delta_2-\frac{23}{15}\alpha\right] \nonumber \\
&&-\frac{1}{20}g^2_1+\frac{4}{5}\Delta_1g_1-\frac{2}{5}\alpha g_1-\frac{4}{5}\alpha^2, \\
\frac{d\Delta_1}{dl}&=&\Delta_1\left[1-\frac{62}{15}\alpha+2\Delta_1+14\Delta_2\right], \\
\frac{d\Delta_2}{dl}&=&\Delta_2\left[1+\frac{22}{15}\alpha+\frac{32}{5}\Delta_2-\frac{6}{5}\Delta_1+\frac{2}{5}g_1-\frac{14}{5}g_2\right] \nonumber \\
&&+\frac{2}{5}\Delta^2_1.
\end{eqnarray}
As in the 2D case both interaction and disorder flow into the strong coupling regime and these equations do not produce a generic quantum critical point. This remains true even if we set $\alpha=0$. In particular, the quantum critical point found for the short range interactions gets ruined by disorder. This is actually not surprising. The Landau theory mandates that the spin-orbit type of disorder mentioned above couples to the rotational symmetry breaking non-magnetic order parameter like a random field, and we have to consider a random field Ising model to understand the stability of the broken symmetry phase. According to the celebrated Imry-Ma argument~\cite{Imry} the rotational symmetry breaking order parameter is destabilized below $d=4$, and at long length scale we should be left with a disorder controlled diffusive metal. Therefore, in the presence of disorder only a magnetically ordered phase can be the most viable broken symmetry phase. However, the competition between the magnetism and the disorder induced diffusive metal is beyond the scope of present weak coupling analysis.  

\section{Conclusions}\label{Sec:VII} 
In this paper we have considered the stability of the parabolic semimetal in two and three spatial dimensions, which are examples of $z=2$ fermionic quantum critical systems. A perturbative renormalization group analysis of the disordered non-interacting model shows that the parabolic fixed point is vanquished by disorder in both 2D and 3D. The resulting disorder controlled phases in 2D and 3D respectively correspond to an Anderson insulator and a diffusive metal. The RG flow equations for the clean interacting system show that the coupling constant for long range Coulomb interaction has a tendency to flow into a NFL fixed point first considered by Abrikosov and Beneslavskii. However, the long range Coulomb interaction also generates short range interaction in the process of coarse graining, which ultimately destabilizes the NFL fixed point, in favor of an excitonic condensate in both 2D and 3D. When both interaction and disorder effects are combined, the rotational (or translational) symmetry breaking non-magnetic phases (e. g., charge density wave and nematic phases) get destroyed according to Imry-Ma argument below $d=4$. For this reason, the time-reversal symmetry breaking order parameters will be most favorable candidates for the broken symmetry phase. For such states (either magnetically ordered or anomalous quantum Hall states in 2D) a sharp quantum phase transition between the disordered controlled metal/insulator and the broken symmetry phase becomes viable. Our calculations suggest that such quantum critical phenomena can be investigated in bilayer graphene (2D) and Pr$_2$Ir$_2$O$_7$ (3D) in future experiments.

\appendix

\section{Fierz identity for three dimensional parabolic semimetal}\label{App:A}

Let us define the interacting Lagrangian for three dimensional parabolic semimetals as
\begin{equation}
L_{int}=g_1 S^2_0 +  g_2 V^2_i + g_3 V_{ij}^2,
\end{equation}
where
\begin{equation}
S_0= \psi^\dagger \psi, \: V_i=\psi^\dagger \Gamma_i \psi, \: V_{ij}= \psi^\dagger \Gamma_{ij} \psi,
\end{equation}
where $\Gamma_{ij}=i \Gamma_i \Gamma_j$ and $j>i$ for $V_{ij}$. Summation over the repeated indices $i,j$ is assumed. The Fierz identity allows one to rewrite each quartic terms as a linear combination of rest according to
\begin{eqnarray}
\left[ \psi^\dagger(x) M \psi(x) \right] \left[ \psi^\dagger(y) N \psi(y) \right]=-\frac{1}{16} \mbox{\bf Tr}\left( M \tilde{\Gamma}_a N \tilde{\Gamma}_b \right) \nonumber \\
  \times  \left[ \psi^\dagger(x) \tilde{\Gamma}_b \psi(y) \right] \: \left[ \psi^\dagger(y) \tilde{\Gamma}_a \psi(x) \right],
\end{eqnarray}
where $\tilde{\Gamma}_a$ with $a=1,2, \cdots, 16$ constitute the basis for all four dimensional Hermitian matrices and $\left(\tilde{\Gamma}_a\right)^{-1}=\tilde{\Gamma}_a$. Let us now define a vector $X^\top=\left(S_0, V_i, V_{ij} \right)$. Above set of constraints can be written as $F\; X=0$, where $F$ is an asymmetric matrix
\begin{equation}
F=\left( \begin{array}{ccc}
5 & 1 & -1 \\
5 & 1 & -1 \\
10 & 2 & -2
\end{array}
\right).
\end{equation}
Rank of $F$ is $1$. Therefore, number of independent four-fermion interaction is $3-1=2$. We can choose $S_0$ and $V_i$ as independent four-fermion interactions, and the remaining interaction $V_{ij}$ can be expressed as linear combination of $S_0$ and $V_i$ as shown in Eq.~(\ref{fierz3d}).

\section*{Acknowledgements} 

We thank I. Herbut, K. Yang and S. Nakatsuji for many fruitful discussions. H.-H. Lai is supported by the National Science Foundation through grant No. DMR-1004545. R. was supported by the start up grant of J. D. Sau from the University of Maryland. P. G. was supported by the NSF Cooperative Agreement No.DMR- 0654118, the State of Florida, and the U. S. Department of Energy.


\begin{thebibliography}{}

\bibitem{Volovik} G. E. Volovik, \emph{Universe in a helium droplet} (Oxford University Press, 2003).

\bibitem{Ghosal} A. Ghosal, P. Goswami, and S. Chakravarty, Phys. Rev. B \textbf{75}, 115123 (2007).  

\bibitem{goswami-chakravarty}  
   P. Goswami, and S. Chakravarty, Phys. Rev. Lett. {\bf 107}, 196803 (2011).

\bibitem{graphene-1} 
    K. S. Novoselov, A. K. Geim, S. V. Morozov, D. Jiang, Y. Zhang, S. V. Dubonos, I. V. Grigorieva, and A. A. Firsov, Science {\bf 306}, 666 (2004).

\bibitem{graphene-2}
   K. S. Novoselov, A. K. Geim, S. V. Morozov, D. Jiang, M. I. Katsnelson, I. V. Grigorieva, S. V. Dubonos, and  A. A. Firsov, Nature {\bf 438}, 197 (2005).

\bibitem{graphene-3}
   Y. Zhang, Y-W. Tan, H. L. Stormer, and P. Kim, Nature {\bf 438}, 201 (2005).
	
\bibitem{Lenoir}B. Lenoir, M. Cassart, J. -P Michenaud, H. Scherrer, and S. Scherrer, J. Phys. Chem. Solids \textbf{57}, 89 (1996), and references therein.	

\bibitem{Dornhaus} R. Dornhaus, G. Nimtz, and B. Schlicht, \emph{Narrow-Gap Semicounductors}, (Springer-Verlag, 1983).

\bibitem{weylexperiment1} 
   S. Borisenko, Q. Gibson, D. Evtushinsky, V. Zabolotnyy, B. Buechner, and R. J. Cava, Phys. Rev. Lett. {\bf 113}, 027603 (2014).
 
\bibitem{weylexperiment2} 
   Z. K. Liu, B. Zhou, Z. J. Wang, H. M. Weng, D. Prabhakaran, S.-K. Mo, Y. Zhang, Z. X. Shen, Z. Fang, X. Dai, Z. Hussain, and Y. L. Chen, Science, {\bf 343}, 864 (2014).

\bibitem{bilayer}
   K. S. Novoselov, E. McCann, S. V. Morozov, V. I. Fal’ko, M. I. Katsnelson, U. Zeitler, D. Jiang, F. Schedin, and A. K. Geim, Nature Physics {\bf 2}, 177 (2006).

\bibitem{alpha-tin-1} S. Grove, and W. Paul, Phys. Rev. Lett. {\bf 11}, 194 (1963).

\bibitem{Pr-1}
   S. Nakatsuji, Y. Machida, Y. Maeno, T. Tayama, T. Sakakibara, J. van Duijn, L. Balicas, J. N. Millican, R. T. Macaluso, and J. Y. Chan, Phys. Rev. Lett. {\bf 96}, 087204 (2006).

\bibitem{Pr-2} 
   Y. Machida, S. Nakatsuji, S. Onoda, T. Tayama, and T. Sakakibara, Nature {\bf 463}, 210 (2010).
	

\bibitem{hassan-cava} 
   S.-Y. Xu, Y. Xia, L. A. Wray, S. Jia, F. Meier, J. H. Dil, J. Osterwalder, B. Slomski, A. Bansil, H. Lin, R. J. Cava,
and M. Z. Hasan, Science {\bf 332}, 560 (2011).

\bibitem{ando} 
   T. Sato, K. Segawa, K. Kosaka, S. Souma, K. Nakayama, K. Eto, T. Minami, Y. Ando, and T. Takahashi, Nat. Phys. {\bf 7}, 840 (2011).

\bibitem{hassan-neupane} 
   M. Brahlek, N. Bansal, N. Koirala, S.-Y. Xu, M. Neupane, C. Liu, M. Z. Hasan, and S. Oh, Phys. Rev. Lett. {\bf 109}, 186403 (2012). 

\bibitem{armitage} 
   L. Wu, M. Brahlek, R. V. Aguilar, A. V. Stier, C. M. Morris, Y. Lubashevsky, L. S. Bilbro, N. Bansal, S. Oh, N. P. Armitage, Nat. Phys. {\bf 9}, 410 (2013) 

\bibitem{TPT-BiT} 
   X. Xi, C. Ma, Z. Liu, Z. Chen, W. Ku, H. Berger, C. Martin, D. B. Tanner, G. L. Carr, Phys. Rev. Lett. {\bf 111}, 155701 (2013).

\bibitem{zhang-model-PRB} 
   C-X. Liu, X-L. Qi, H-J. Zhang, X. Dai, Z. Fang, and S-C. Zhang, Phys. Rev. B {\bf 82}, 045122 (2010).
	
\bibitem{CYan} C. Yan, J. Liu, Y. Zang, J. Wang, Z. Wang, P. Wang, Z-D. Zhang, L. Wang, X. Ma, S. Ji, K. He, L. Fu, W. Duan, Q-K. Xue, and X. Chen, Phys. Rev. Lett. \textbf{112}, 186801 (2014).	
	

\bibitem{luttinger}  
   J. M. Luttinger, Phys. Rev. {\bf 102}, 1030 (1956).

\bibitem{murakami-zhang-nagaosa} 
   S. Murakami, S-C. Zhang, and N. Nagaosa,  Phys. Rev. B {\bf 69}, 235206 (2004).

\bibitem{kveshchenko} 
   D. V. Khveshchenko, Phys. Rev. Lett. {\bf 87}, 246802 (2001); H. Leal and D. V. Khveshchenko, Nucl. Phys. {\bf B687}, 323 (2004).

\bibitem{HJR} 
   I. F. Herbut, V. Juri\v ci\' c, and B. Roy, Phys. Rev. B {\bf 79}, 085116 (2009); I. Herbut, B. Roy, Phys. Rev. B {\bf 82}, 035429 (2010).

\bibitem{drut} 
   J. E. Drut and T. A. L\" ahde, Phys. Rev. Lett. {\bf 102}, 026802 (2009); Phys. Rev. B {\bf 79}, 165425 (2009).

\bibitem{aji} 
   H. Wei, S-P. Chao, and V. Aji, Phys. Rev. Lett. {\bf 109}, 196403 (2012).

\bibitem{nandkishore-weyl} 
   J. Maciejko, and R. Nandkishore, Phys. Rev. B {\bf 90} 035126 (2014).

\bibitem{abrikosov-beneslavski}  
   A. A. Abrikosov and S. D. Beneslavskii, Sov. Phys. JETP {\bf 32}, 699 (1971).

\bibitem{abrikosov}  
   A. A. Abrikosov, Sov. Phys. JETP {\bf 39}, 709 (1974).

\bibitem{balents-kim} 
   E-G. Moon, C. Xu, Y. B. Kim, L. Balents, Phys. Rev. Lett. {\bf 111}, 206401 (2013).

\bibitem{serrington-kohn} 
   D. Sherrington and W. Kohn, Rev. Mod. Phys. {\bf 40}, 767 (1968), and references therein. 

\bibitem{herbut-jansen} 
   I. F. Herbut, and L. Janssen, Phys. Rev. Lett. {\bf 113}, 106401 (2014).

\bibitem{vafek} 
   O. Vafek and K. Yang, Phys. Rev B {\bf 81}, 041401(R) (2010); R. E. Throckmorton, O. Vafek, \emph{ibid.} {\bf 86}, 115447 (2012).

\bibitem{nandkishore} 
   R. Nandkishore and L. Levitov, Phys. Rev. Lett. {\bf 104}, 156803 (2010); Phys. Rev. B {\bf 82}, 115124 (2010).

\bibitem{aleiner} 
   Y. Lemonik, I. L. Aleiner, C. Toke, and V. I. Fal’ko, Phys. Rev. B {\bf 82}, 201408(R); Y. Lemonik, I. L. Aleiner, and V. I. Fal'ko, \emph{ibid.} {\bf 85}, 245451 (2012).

\bibitem{fanzhang} 
   F. Zhang, H. Min, M. Polini, and A. H. MacDonald, Phys. Rev. B {\bf 81}, 041402 (2010); J. Jung, F. Zhang, A. H. MacDonald, \emph{ibid.} {\bf 83}, 115408 (2011). 

\bibitem{bitan-BLG}
   B. Roy, Phys. Rev. B {\bf 89}, 201401(R) (2014); {\it ibid} {\bf 88}, 075415 (2013).

\bibitem{BLG-exp1} 
   A. S. Mayorov, D. C. Elias, M. Mucha-Kruczynski, R. V. Gorbachev, T. Tudorovskiy, A. Zhukov, S. V. Morozov, M. I. Katsnelson, V. I. Fal'ko, A. K. Geim, K. S. Novoselov, Science, {\bf 333}, 860 (2011).

\bibitem{BLG-exp2} R. T. Weitz, M. T. Allen, B. E. Feldman, J. Martin, A. Jacoby, Science {\bf 330}, 812 (2010).
 
\bibitem{BLG-exp3} 
   J. Velasco Jr., L. Jing, W. Bao, Y. Lee, P. Kratz, V. Aji, M. Bockrath, C.N. Lau, C. Varma, R. Stillwell, D. Smirnov, Fan Zhang, J. Jung, A.H. MacDonald, Nat. Nano. {\bf 7}, 156 (2012). 
 
\bibitem{BLG-exp4} 
   F. Freitag, M. Weiss, R. Maurand, J. Trbovic, and C. Schonenberger, Phys. Rev. B {\bf 87}, 161402 (R) (2013); Solid State Commun. {\bf 152}, 2053 (2012).

\bibitem{sun-fradkin-kivelson}
   K. Sun, H. Yao, E. Fradkin, and S. Kivelson, Phys. Rev. Lett. {\bf 103}, 046811 (2011).

\bibitem{2d-diorderdirac}
    I. L. Aleiner, and K. B. Efetov, Phys. Rev. Lett. {\bf 97}, 236801 (2006);  M. S. Foster, and I. L. Aleiner, Phys. Rev. B {\bf 77}, 195413 (2008).

\bibitem{fradkin}
   E. Fradkin, Phys. Rev. B {\bf 33}, 3263 (1985).

\bibitem{roy-dassarma}
   B. Roy, and S. Das Sarma, arxiv:1407.7026.
	
\bibitem{RoyHerbutJuricic} B. Roy, V. Juricic, I. F. Herbut, Phys. Rev. B \textbf{87}, 041401 (R) (2013).
	
\bibitem{Mirlin_nonlinearsigma}
   P. M. Ostrovky, I. V. Gornyi, and A. D. Mirlin, Phys. Rev. Lett. {\bf 98}, 256801 (2007).

\bibitem{Ryu_nonlinearsigma}
   S. Ryu, C. Mudry, H. Obuse, and A. Furusaki, Phys. Rev. Lett. {\bf 99}, 116601 (2007).
	
\bibitem{Barlas} Y. Barlas and K. Yang, Phys. Rev. B \textbf{80}, 161408(R) (2009).  	
	
\bibitem{Pruisken} H. Levine, S. B. Libby, and A. M. M. Pruisken, Phys. Rev.
Lett. \textbf{51}, 1915 (1983).	

\bibitem{Imry} Y. Imry and S-K. Ma, Phys. Rev. Lett. \textbf{35}, 1399 (1975).
	

\end{thebibliography}
\end{document}